\documentclass[aip,pop,reprint,superscriptaddress]{revtex4-1}
\usepackage{dcolumn}
\usepackage{bm}
\usepackage{amssymb,amsmath,amsthm,epsfig}
\usepackage{sidecap}

\usepackage{natbib}

\def\apj{ApJ.}
\def\solphys{Sol. Phys.}
\def\aap{A\&A}

\def\jgr{J. Geophys. Res.}

\def\apjl{Astrophys. J. Lett}

\begin{document}

\title{Kelvin-Helmholtz instability in a current-vortex sheet at a 3D magnetic null}
\author{P. F. Wyper}
\email{app09pfw@sheffield.ac.uk}
\affiliation{School of Mathematics and Statistics, University of Sheffield, S3 7RH, UK}
\author{D. I. Pontin}
\email{dpontin@maths.dundee.ac.uk}
\affiliation{Division of Mathematics, University of Dundee, DD1 4HN, UK}

\date{\today}

\begin{abstract}
We report here, for the first time, an observed instability of a Kelvin-Helmholtz (KH) nature occurring in a fully three-dimensional (3D) current-vortex sheet at the fan plane of a 3D magnetic null point. The current-vortex layer forms self-consistently in response to foot point driving around the spine lines of the null. The layer first becomes unstable at an intermediate distance from the null point, with the instability being characterized by a rippling of the fan surface and a filamentation of the current density and vorticity in the shear layer. Owing to the 3D geometry of the shear layer, a branching of the current filaments and vortices is observed. The instability results in a mixing of plasma between the two topologically distinct regions of magnetic flux on either side of the fan separatrix surface, as flux is reconnected across this surface. We make a preliminary investigation of the scaling of the system with the dissipation parameters. Our results indicate that the fan plane separatrix surface is an ideal candidate for the formation of current-vortex sheets in complex magnetic fields and, therefore, the enhanced heating and connectivity change associated with the instabilities of such layers.
\end{abstract}


\maketitle

\section{Introduction}
In recent years three-dimensional (3D) magnetic null points (points in space where the magnetic field strength is zero) have become increasingly recognized as important sites for energy release and magnetic topology change through magnetic reconnection. They are found in abundance in the lower solar atmosphere \citep{Regnier2008,Longcope2009} whilst during active times of the solar cycle they have been inferred to be involved in solar jets \citep{Pariat2009,Liu2011}, flux emergence \citep{Torok2009}, flare brightening \citep{Masson2009} and magnetic breakout \citep{Lynch2008}. In the Earth's magnetosphere their existence has been confirmed in the reconnecting magnetotail current sheet through {\it{in situ}} measurements  \citep[e.g.][]{Xiao2006} whilst clusters of nulls have also been seen in the polar cusp regions during global magnetospheric simulations  \citep{Dorelli2007}.

The magnetic field near to a 3D null can be characterized by two main topological structures. The fan plane: a surface of field lines which emanate from (approach toward) the null and the spine lines: two field lines which approach toward (recede away from) the null. Reconnection at 3D nulls can take place in several ways. Twisting motions about the spine/fan create a current sheet along the fan/spine leading to a rotational slippage known as {\it torsional fan/spine reconnection}, respectively \citep{PriestPontin2009}. Shearing motions of the spine or fan in an incompressible plasma give rise to current accumulation at the fan and spine respectively \citep{PriestTitov1996,CraigFabling1996} known as {\it fan} and {\it spine reconnection}. However, when the incompressibility assumption is relaxed, the plasma pressure gradient is too weak within the planar {(spine or fan)} current sheet to oppose the collapse of the field around the null (driven by the Lorentz force), and a current sheet forms at an angle to the spine and fan within which {\it spine-fan reconnection} takes place \citep{PriestPontin2009,Galsgaard2011b,PontinBhatt2007,PontinBhatt2007b}. Lastly, it has recently been suggested \cite{Jorge2012} on the basis of an infinite-time singularity that a new regime exists for initially asymmetric nulls where shear builds aligned to the minor fan plane axis, resulting in a current sheet with a strong component parallel to the spine. 

In the cases of externally driven fan and torsional fan reconnection the current layer that forms at the fan separatrix surface includes both a sheared magnetic and velocity field component. Studies of both regimes have until now focused purely on the formation of these smooth current layers. However, such a shear layer configuration (known as a current-vortex sheet) is know to be unstable to shear flow and resistive instabilities leading to fragmentation of the current layer and multiple reconnection sites. These instabilities have perhaps not been observed in previous 3D null point simulations owing to the requirement of rather large resistivity and viscosity and the relatively short time scales over which the boundary driving velocities were imposed. We report here, for the first time, an observed instability of a Kelvin-Helmholtz (KH) nature occurring in a fully three-dimensional current-vortex sheet at the fan plane of a 3D null point.

The KH instability has attracted significant attention for a number of years, both in the hydrodynamic limit and in the presence of a magnetic field \citep[e.g.][]{Chandrasekhar1961,Miura1982}. A uniform magnetic field component aligned with the shear flow is known to be a stabilizing influence, due to the associated magnetic tension force. On the other hand, a magnetic field component perpendicular to the shear layer (or `transverse field') does not affect the stability, but modifies the growth rate of a given mode \citep[e.g.][]{Miura1982}. In our simulations, the instability takes place at the fan plane of a 3D null, and there exists a strong transverse magnetic field component (the radial field $B_r$ associated with the potential field defining the null), as well as a sheared in-plane component that reverses sign at the same location as the flow (the azimuthal field $B_\theta$ associated with the current layer that forms at the fan in response to the boundary driving). The stability of such a current-vortex sheet -- where the magnetic and velocity shear layers coincide -- has been previously studied in 2- and 2.5-dimensions \cite{Einaudi1986,Dahlburg1997,Keppens1999,Dahlburg2002}. 
Einaudi {\it et al.} \cite{Einaudi1986} showed that, in the incompressible limit, a transition between a tearing-like regime and a KH-like regime occurs when $\Lambda=\left(\frac{L_{b}}{L_{v}}\right)\left(\frac{\Delta v}{c_{A}}\right)^{2/3} = 1$. Here $L_{b}$ and $L_{v}$ are the widths of the magnetic and velocity shear layers, $\Delta v$ is the velocity difference across the layer and $c_{A}$ the Alfv\'{e}n speed far from the layer. When $\Lambda<1$, a tearing unstable regime is found as the magnetic shear strongly outweighs the velocity shear. Conversely, when $\Lambda>1$ the velocity shear dominates the layer, and the linear phase of the instability is ideal, and of a KH nature. This transition has also been shown to hold in weakly compressible \citep{Dahlburg2000} and viscous plasmas \citep{Einaudi1989}. Dahlburg {\it et al.} \cite{Dahlburg1997} argue that even when $\Lambda>1$, the presence of the magnetic shear fundamentally alters the nature of the KH-type instability by allowing magnetic reconnection to become important, and that therefore the instabilities of the current-vortex sheet should not be considered as a simple mix of tearing and KH modes.
 
The magnetohydrodynamic (MHD) KH instability is important in a broad range of plasma environments, both terrestrial and astrophysical. There are numerous observations of KH signatures on planetary magnetopauses \cite[e.g.][]{hasegawa2004,masters2010,sundberg2012}. The KH instability is also crucial in the disruption of astrophysical jets, emanating for example from young stellar objects and active galactic nuclei \citep{ferrari1998}.
In the solar corona there are recent direct observations of KH instabilities in regions of strong flow shear associated with eruptions \citep{Foullon2011,Ofman2011}. Our work highlights that, in addition to these fast flows, magnetic separatrices are prime locations for the formation of current-vortex sheets in magnetic fields of complex topology such as the solar corona. These separatrix surfaces have already been proposed as preferential sites of plasma heating in the corona \citep[e.g.][]{priest2005}, which could be significantly enhanced by instabilities such as that studied herein.

In this work we will investigate the self consistent formation and stability of the current-vortex sheet created by twisting motions around the spine foot points of a linear, rotationally symmetric 3D magnetic null. We restrict our focus to the KH unstable regime ($\Lambda>1$), leaving the tearing-type regime to a future study. It should be noted, however, that in the tearing unstable regime the growth of the instability would be expected to be slow as a velocity shear flow is known to damp the tearing mode growth rate \citep{Chen1990}. We also note that here the geometry of the problem is much more complex than that of the studies discussed above. In particular, the widths of the shear layers ($L_B$ and $L_v$) vary along the transverse (radial) direction, and are set in a self-consistent manner by a balance between the driving flow and the dissipation in the system, rather than being fixed by the initial conditions.

The investigation is structured as follows: in Section 2 the numerical setup is discussed. In Sections 3 and 4 we investigate, respectively, the formation of the current-vortex layer, and its breakup. In Sections 5 and 6 we discuss our findings and present our conclusions. 

\section{Numerical Setup}
The investigation was carried out numerically by solving the compressible MHD equations in the following form
\begin{eqnarray}
\frac{\partial \mathbf{B}}{\partial t} &=& \mathbf{\nabla}\times (\mathbf{v}\times\mathbf{B}) +\eta \mathbf{\nabla}^{2}\mathbf{B}\\
\frac{\partial(\rho \mathbf{v})}{\partial t} &=& -\mathbf{\nabla}\cdot(\rho \mathbf{v}\mathbf{v}) -\mathbf{\nabla}p  +\mathbf{j}\times\mathbf{B} \nonumber\\
&\quad&\;+\quad \mu\left(\mathbf{\nabla}^{2}\mathbf{v}+\frac{1}{3}\mathbf{\nabla}(\mathbf{\nabla}\cdot\mathbf{v})\right)\\
\frac{\partial e}{\partial t}&=& -\mathbf{\nabla}\cdot(e\mathbf{v})-p\mathbf{\nabla}\cdot\mathbf{v}+\eta j^2 +Q_{visc}\\
\frac{\partial \rho}{\partial t}&=&-\nabla\cdot(\rho\mathbf{v})\\
\mathbf{j}&=&\mathbf{\nabla}\times \mathbf{B}/\mu_{0}\\
\mathbf{\nabla}\cdot\mathbf{B}&=&0
\end{eqnarray}
with velocity $\mathbf{v}$, magnetic field $\mathbf{B}$, density $\rho$, thermal energy $e$, gas pressure $p=(\gamma-1)e$, magnetic diffusivity $\eta$ and dynamic viscosity $\mu$. The viscous heating term is 
$$
Q_{visc}=\mu\left(\frac{{\partial v_i}}{\partial x_j}\frac{{\partial v_i}}{\partial x_j}+\frac{{\partial v_j}}{\partial x_i}\frac{{\partial v_i}}{\partial x_j}-\frac{2}{3}(\nabla \cdot {\bf v})^2\right),
$$  
using the convention of summation over repeated indices. In most of our simulations we use prescribed, spatially-uniform $\eta$ and $\mu$.

The equations are solved on staggered grids with a 6th order method applied to find the partial derivatives, a 5th order method used for interpolation, and a 3rd order predictor-corrector method used to advance the code in time. In the simulations described herein, the velocity components perpendicular to the boundaries are set to zero, implying no mass flux through the boundaries. On the $x$-boundaries, we restrict parallel velocities to be non-zero only in the two regions where a prescribed stressing velocity flow is imposed (see below). Near the y- and z- boundaries a strong damping region is in place (discussed below) so that in practice parallel velocities on these boundaries are negligible. Further information relating to the code can be found in Ref.~\onlinecite{Galsgaard1997} and on \url{http://www.astro.ku.dk/~kg/}.

The equations are non-dimensionalised by setting the magnetic permeability $\mu_{0}=1$ and the gas constant equal to the mean molecular weight. This results in one time unit representing the travel time of an Alfv\'{e}n wave over a unit distance through a plasma with unit density and unit magnetic field ($\rho=1, |\mathbf{B}|=1$). This also means that magnetic diffusivity is equal to magnetic resistivity and takes the form of an inverse magnetic Reynolds number $\eta=\frac{\eta_{dim}}{LV_{0}}=Re_{m}^{-1}$ and the kinematic viscosity ($\nu=\mu/\rho$) takes the form of an inverse plasma Reynolds number $\nu=\frac{\nu_{dim}}{LV_{0}}=Re^{-1}$, where $L$ and $V_{0}$ are some typical length scale and velocity. Note that for simplicity we have neglected the effects of thermal conduction in the energy equation. This means that the temperature is not equilibrated along field lines. If conduction were included we would expect to see a modification of the plasma properties in the current-vortex layer that forms -- see the discussion later.

To investigate the properties of the current-vortex sheet created through twisting motions around the spine we start with a linear, rotationally symmetric null point with magnetic field $\mathbf{B}=B_{0}(-2x,y,z)$ in the center of a numerical box of size $\pm[0.25,3.5,3.5]$. The grid is stretched to include more points near to the spine and fan of the null to improve the resolution of structures there. The plasma is assumed to be an ideal gas with $\gamma=5/3$ and is initially at rest with density $\rho=1$ and thermal energy $e=5\beta^{*}/2$. Here $\beta^{*}$ is a parameter that controls the plasma-$\beta$, the ratio of plasma pressure to magnetic pressure: $P/(B^{2}/2\mu_{0})=10\mu_{0}\beta^{*}/3 B^{2}$. We set $\beta^{*}=0.5$ and $B_{0}=1$ in all the simulations. By setting $\rho=1$ initially, the kinematic viscosity becomes the same as the dynamic viscosity ($\nu=\mu/\rho=\mu$) and so the magnetic Prandtl number may be found from $Pr=\mu/\eta$.

Because of the cylindrical symmetry of our system it is convenient to define a new cylindrical coordinate system
\begin{equation}
\hat{\mathbf{r}}=r\cos{\theta} \hat{\mathbf{y}} + r\sin{\theta} \hat{\mathbf{z}}, \quad \hat{\boldsymbol{\theta}}=-r\sin{\theta} \hat{\mathbf{y}} + r\cos{\theta} \hat{\mathbf{z}},\quad \hat{\mathbf{x}}=\hat{\mathbf{x}} \nonumber
\end{equation}
where $r=\sqrt{y^2+z^2}$ and $\theta=\tan^{-1}\left(\frac{z}{y}\right)$. The $\hat{\mathbf{r}}$ and $\hat{\boldsymbol{\theta}}$ directions are referred to as the radial direction and azimuthal directions respectively. In these coordinates the magnetic field becomes $\mathbf{B}=r\hat{\bf r}-2x\hat{\bf x}$. Rotational driving on the boundaries is applied in opposite senses around each spine foot point of the form
\begin{equation}
v_{\theta}(x=\mp 0.25) = \pm V_{0}(t) r \left(1+\tanh\left((1-36r^{2})\right)\right) 
\end{equation}
where
\begin{equation}
V_{0}(t)=v_{0} \tanh^{2}\left(\frac{t}{\tau}\right).
\end{equation}
We choose $\tau=0.25$ to smoothly ramp the driving up from zero. It has been previously noted that a near-discontinuous increase in driving generates fast waves, in addition to the main torsional Alfv\'{e}n wave, which focus on to the null \citep{Galsgaard2003}. We ramp the driving up slowly to avoid this additional complication to the evolution.

Lastly, as we have a cylindrically symmetric system, we impose a cylindrical damping region beyond $r=2.8$ which removes momentum in a way which increases linearly with radius ($r$), and limits the reflection of disturbances from the $y$ and $z$ boundaries. Thus, if we consider the side boundary to be the edge of the damping region ($r=2.8$) this boundary can be considered quasi-open.

\begin{table}[ht]
\caption{Summary of simulations} 
\centering 
\begin{tabular}{c c c c c c c} 
\hline\hline 
Case & $v_{0}$ & $\eta$ & $\mu$ & $Pr=\nu/\eta$ & Resolution & Stable? \\ [0.5ex] 
\hline 
1 & 0.25 & $5\times 10^{-4}$ & $1\times 10^{-4}$ & 0.2 & $160^{3}$ & Yes\\ 
2 & 0.5 & $5\times 10^{-4}$ & $1\times 10^{-4}$ & 0.2 & $160^{3}$  & Yes\\
3 & 1.0 & $5\times 10^{-4}$ & $1\times 10^{-4}$ & 0.2 & $160^{3}$  & Yes\\
4 & 0.5 & $2\times 10^{-4}$ & $1\times 10^{-4}$ & 0.5 & $160^{3}$  & Yes\\
5 & 0.5 & $1\times 10^{-3}$ & $1\times 10^{-4}$ & 0.1 & $160^{3}$  & Yes\\ 
6 & 0.5 & $5\times 10^{-4}$ & $1\times 10^{-3}$ & 2.0 & $160^{3}$  & Yes\\
7 & 0.5 & $5\times 10^{-4}$ & $5\times 10^{-4}$ & 1.0 & $160^{3}$  & Yes\\
8 & 0.5 & $2\times 10^{-4}$ & $1\times 10^{-5}$ & 0.05 & $320^{3}$ & No\\
9 & 0.5 & $5\times 10^{-4}$ & $1\times 10^{-5}$ & 0.02 & $320^{3}$ & No\\
10 & 0.5 & $1\times 10^{-3}$ & $1\times 10^{-5}$ & 0.01 & $320^{3}$ & No\\
11 & 0.5 & $2\times 10^{-4}$ & numerical & - & $320^{3}$ & No\\
12 & 0.5 & $5\times 10^{-4}$ & numerical & - & $320^{3}$ & No\\
13 & 0.5 & $1\times 10^{-3}$ & numerical & - & $320^{3}$ & No\\[1ex] 
\hline 
\end{tabular}
\label{table:runs} 
\end{table}

\begin{figure}
\centering
\includegraphics{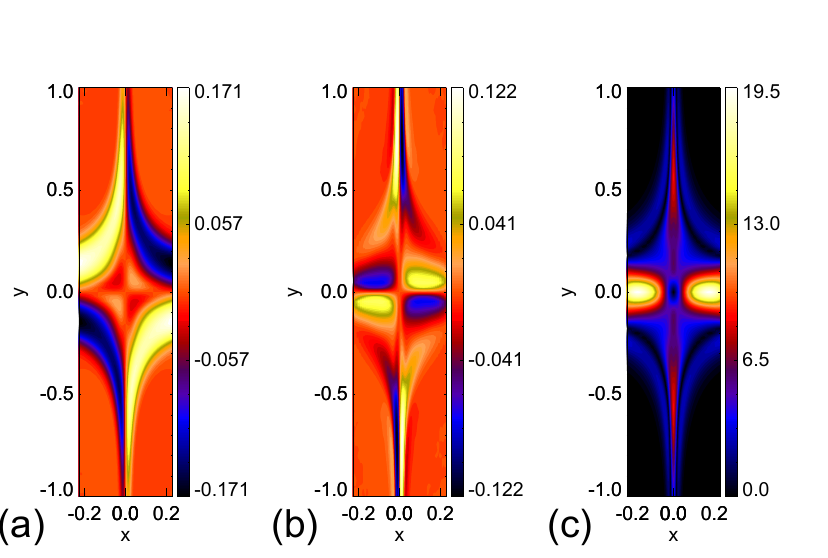}
\caption{Plots viewed at $t=2.4$ in the $z=0$ plane for case 3. The null is at the origin, the spine along $y=0$ and the fan at $x=0$. (a) The velocity out of the plane (note the counter flow regions near the null), (b) the Lorentz force out of the plane and (c) the current density.}
\label{fig:counterflow}
\end{figure}

\section{Formation of the current-vortex sheet}
\subsection{Qualitative Behavior}
We first discuss the formation of the current-vortex sheet, and its structure. Several previous investigations have observed the formation of a current layer focused on the fan plane in response to rotational driving motions \citep{Galsgaard2003,PontinGalsgaard2007,mclaughlin2008,AlHachami2011}. The early stages of the simulations proceed as follows: once the driving begins a torsional Alfv\'{e}n wave is launched from each boundary which spreads out as it follows the hyperbolic shape of the field toward the null. The current in the wave front increases as the length scale (perpendicular to the fan) decreases. Once both wave fronts get close to the fan the current diffuses into the fan plane itself creating a strong current layer. Of the studies cited above, only Galsgaard {\it et al.} \cite{Galsgaard2003} maintained the driving for long enough to see the appearance of counter rotating (i.e. against the direction of the driver) flow regions near the null. However, the focus of that investigation was on the wave dynamics so this was not deeply investigated and no physical reason was put forward for the appearance of these flows. As these counter flows become important for the stability of the current-vortex sheet, we now investigate their properties.

\begin{figure}
\centering
\includegraphics{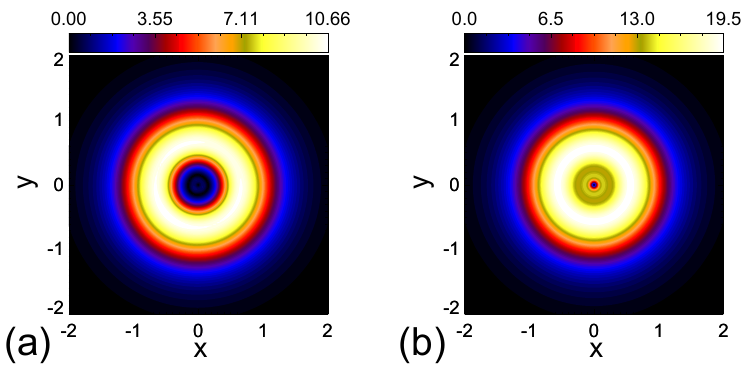}
\caption{Vorticity density (a) and current density (b) in the fan plane ($x=0$) at $t=2.4$ for case 3.}
\label{fig:vortexhole}
\end{figure}

\begin{figure}
\centering
\includegraphics{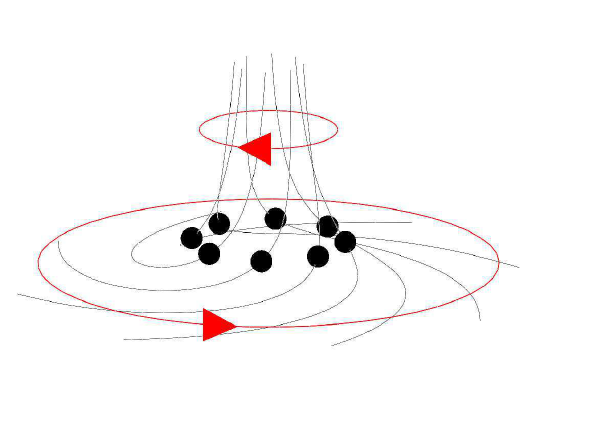}
\caption{Sketch showing the different magnetic tension force in different regions close to the spine and close to the fan. Between the current concentrations at the fan and spine, the field lines pass through an ideal region where they are effectively anchored in the flow (black dots). In the two non-ideal regions the magnetic field lines try to straighten (red arrows) due to the tension force.}
\label{fig:diagrams}
\end{figure}

\subsection{The Counter flow}
\label{sec:counter}
To investigate these flows a series of simulations were performed at a resolution of $160^3$ for various plasma and boundary driving parameters to observe the early dynamics of the system (see table \ref{table:runs}, cases 1 to 7 for details). The viscosity in these cases is chosen to be large enough that the current-vortex sheet remains stable, allowing us to focus on the formation of the counter flow regions. Figure \ref{fig:counterflow} shows an example of the counter rotational flow that begins to form around $t=1.8$ for case 3 (see table \ref{table:runs}). These counter rotating regions locally inhibit the velocity shear across the fan plane and produce what we term a `hole' in the vorticity layer around the null (Fig. \ref{fig:vortexhole}(a)). Despite the drop in velocity shear in this region there still exists a strong shear in the magnetic field across the fan as can be seen by the strength of the current (Fig. \ref{fig:vortexhole}(b)).

The formation of the vortex hole arises from the interplay of forces in the vicinity of the null. Through symmetry considerations the magnetic tension must be the dominant force driving the formation of these flows. Comparing the Lorentz force (identical to the magnetic tension in the observed plane) and velocity plots in Fig. \ref{fig:counterflow}, a clear correlation between the counter flows and the tension force associated with the two regions of strong current along the spine and fan (Fig. \ref{fig:counterflow}(c)) is indeed evident. 
The reason for the development of such a pattern in the tension force is not clear but is likely linked to the manner in which field line slippage occurs within the two current regions. Certainly, as the tension force is opposite in each current concentration, it appears that a typical magnetic field line can be considered as being effectively anchored in the ideal region between the current concentrations, with this point acting as a pivot about which the field line wants to straighten -- see Fig. \ref{fig:diagrams}. This suggests that the field lines around the spine have become partially detached from the driving boundary as a result of the connection change within the strong current concentration near the spine. Were these foot points still strongly attached to the field in the volume, then the magnetic field would be expected to straighten around the foot points themselves (giving a unidirectional tension force between the foot points and the fan plane). In any case, it is clear that the current concentration along the spine has a profound effect on the plasma dynamics near the fan plane.

We find that with stronger driving (cases 1 to 3) or for lower values of resistivity (cases 2, 4 and 5) the current build-up near the spine (and therefore the magnetic tension force in this region) increases. Thus, the strength of counter flows, and therefore the size of the `hole' in the vorticity, also increase. In addition, when $\eta$ is reduced the thinning of the fan current layer combined with the hyperbolic shape of the magnetic field also widen the vorticity hole. As we might expect, a reduction in viscosity (cases 6 and 7) also increases the strength of the counter flows by reducing drag between the shear layers. Lastly, we note that the strength of the current regions near the driving boundaries are strongly dependent upon the chosen driving profile and that other choices of rotational driving will result in different degrees of counter flow. As we shall see later, the counter flow region plays a key role in determining the initial region of instability.

\begin{figure}
\centering
\includegraphics{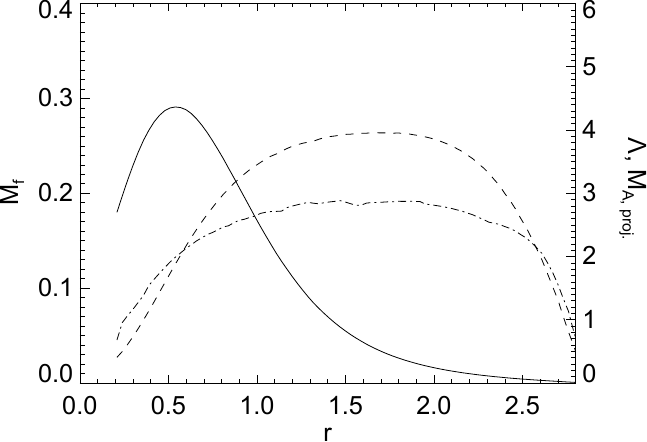}
\caption{Azimuthally symmetric shear layer quantities for case 9 at $t=5$. Solid line: the fast mode Mach number ($M_{f}=\Delta v/\sqrt{c_{s}^2+c_{A}^2}$); dashed line: the projected Alfv\'{e}n Mach number ($M_{A,\; proj.}=\Delta v\sqrt{\rho \mu_{0}}/\Delta B$) and dot-dashed line: $\Lambda$.}
\label{fig:layer}
\end{figure}

\section{KH Instability of the Current-Vortex Sheet}
\subsection{Properties of the shear layer}
Having discussed the formation of the current-vortex sheet at the fan surface, we proceed to explore its stability to the KH-type instability. Two sets of simulations were performed at $320^3$ grid resolution, see table \ref{table:runs}. For one set (cases 8 to 10) $\mu$ was set to zero and therefore viscosity is handled through numerical diffusion. This gives the least damping of any KH fluctuations for the chosen resolution and provides a benchmark against the second set (cases 11 to 13) where $\mu=1\times 10^{-5}$. 

Within the fan plane current sheet the velocity shear is associated with a variation of $v_\theta$ with $x$. Relative to this flow the magnetic field has two components: a strong transverse guide field component $B_r$ associated with the initial potential null point field, and a sheared component $B_\theta$ parallel to the plasma flow (associated with the current in the layer formed in response to the boundary driving). Related to these two components we can define two radially varying but rotationally symmetric Mach numbers:
\begin{equation}
M_{f}=\frac{\Delta v}{\sqrt{c_{s}^{2}+c_{A}^{2}}} \quad \& \quad M_{A,\; proj.}=\frac{\Delta v\sqrt{\rho \mu_{0}}}{\Delta B},
\end{equation}
where $\Delta v$ and $\Delta B$ are the total velocity and magnetic shear across the layer and $c_{s}$ and $c_{A}$ are the sound and Alfv\'{e}n speeds, all of these quantities being dependent on $r$. $M_{f}$ is the fast mode Mach number related to the velocity shear and $M_{A,\; proj.}$ is the projected Alfv\'{e}n Mach number associated with the sheared magnetic field component. For the KH instability, in the case of a constant perpendicular guide field, the instability is linearly stabilized at all wavenumbers when $M_{f}>2$ (Ref.~\onlinecite{Miura1982}). In addition, the transition from a tearing-type to a KH-type instability should occur around $\Lambda=\left(\frac{L_{b}}{L_{v}}\right)M_{A,\; proj.}^{2/3} = 1$ (Ref.~\onlinecite{Einaudi1986}). Therefore, to excite the KH instability we choose the driving and plasma parameters so that the layer is super-Alfv\'{e}nic (in a projected sense) but sub-magentosonic. Figure \ref{fig:layer} shows that these conditions are satisfied at almost all values of $r$ for case 9 just before any instability arises. We note that the reduction in $\Lambda$ beyond $r=2.0$ is in part due to the boundary damping at $r=2.8$, however the instability begins well away from this edge and so the boundaries do not affect the main evolution.

\begin{SCfigure*}
\centering
\includegraphics{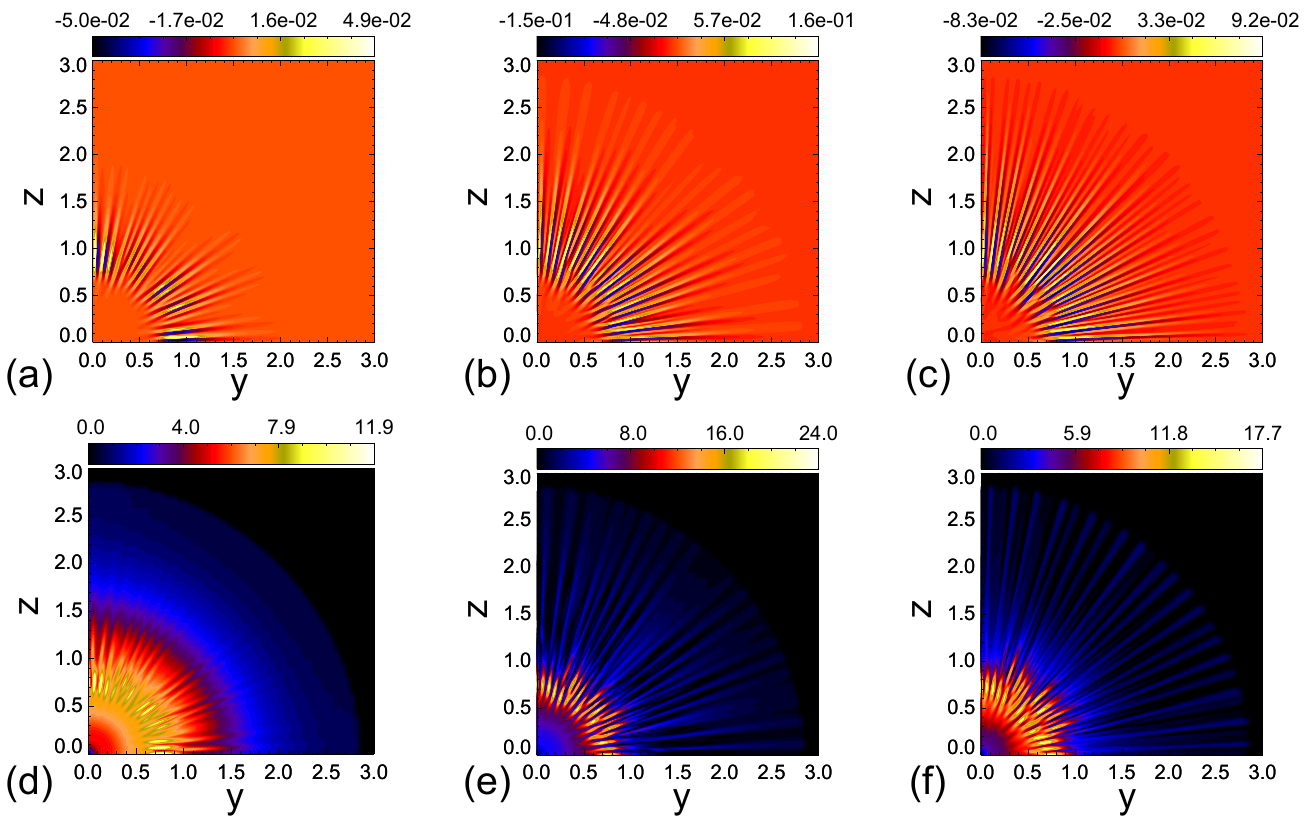}
\caption{(a-c) The velocity out of the plane ($v_{x}(x=0)$). (d-f) The current density in the plane ($|\mathbf{J}|(x=0)$). The contours are scaled to the maximum in each frame. For times $t=7.0$ (a,d), $8.0$ (b,e) and $9.0$ (c,f), for simulation 9.}
\label{fig:rolltop}
\end{SCfigure*}

\begin{figure*}
\centering
\includegraphics{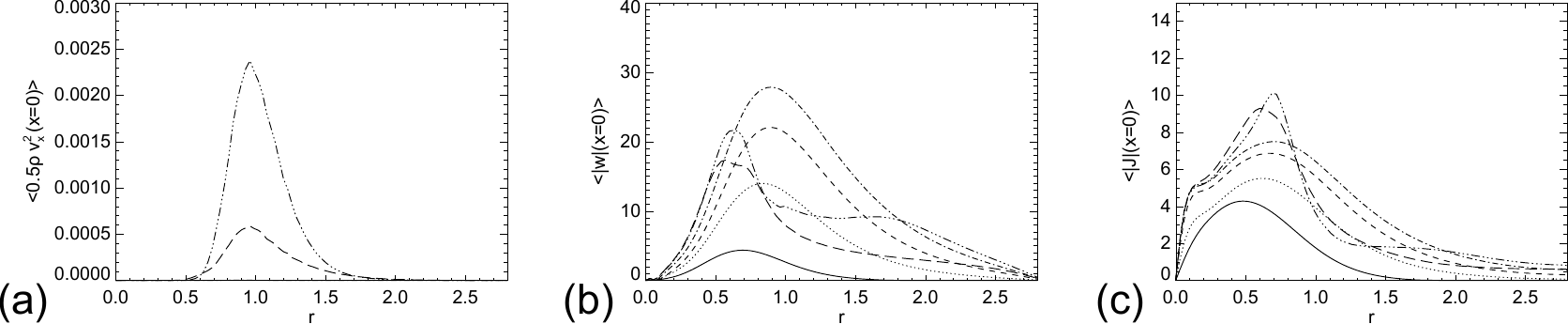}
\caption{Azimuthally averaged quantities plotted as a function $r$ in the $x=0$ plane (the initial position of the fan plane) at different $t$. (a) The average perpendicular kinetic energy ($<\rho v_{x}^{2}(x=0)>$), (b) the average vorticity density ($<|\mathbf{w}|(x=0)>$) and (c) the average current density ($<|\mathbf{J}|(x=0)>$), where $<..>$ denotes an average over the azimuthal angle $\theta=\tan^{-1}(z/y)$. Solid line: $t=2.0$, dotted: $t=3.5$, dashed: $t=5.0$, dot-dashed: $t=6.5$, triple-dot-dashed: $t=8.0$, long dash: $t=9.5$.}
\label{fig:avr}
\end{figure*}

\begin{SCfigure*}
\centering
\includegraphics{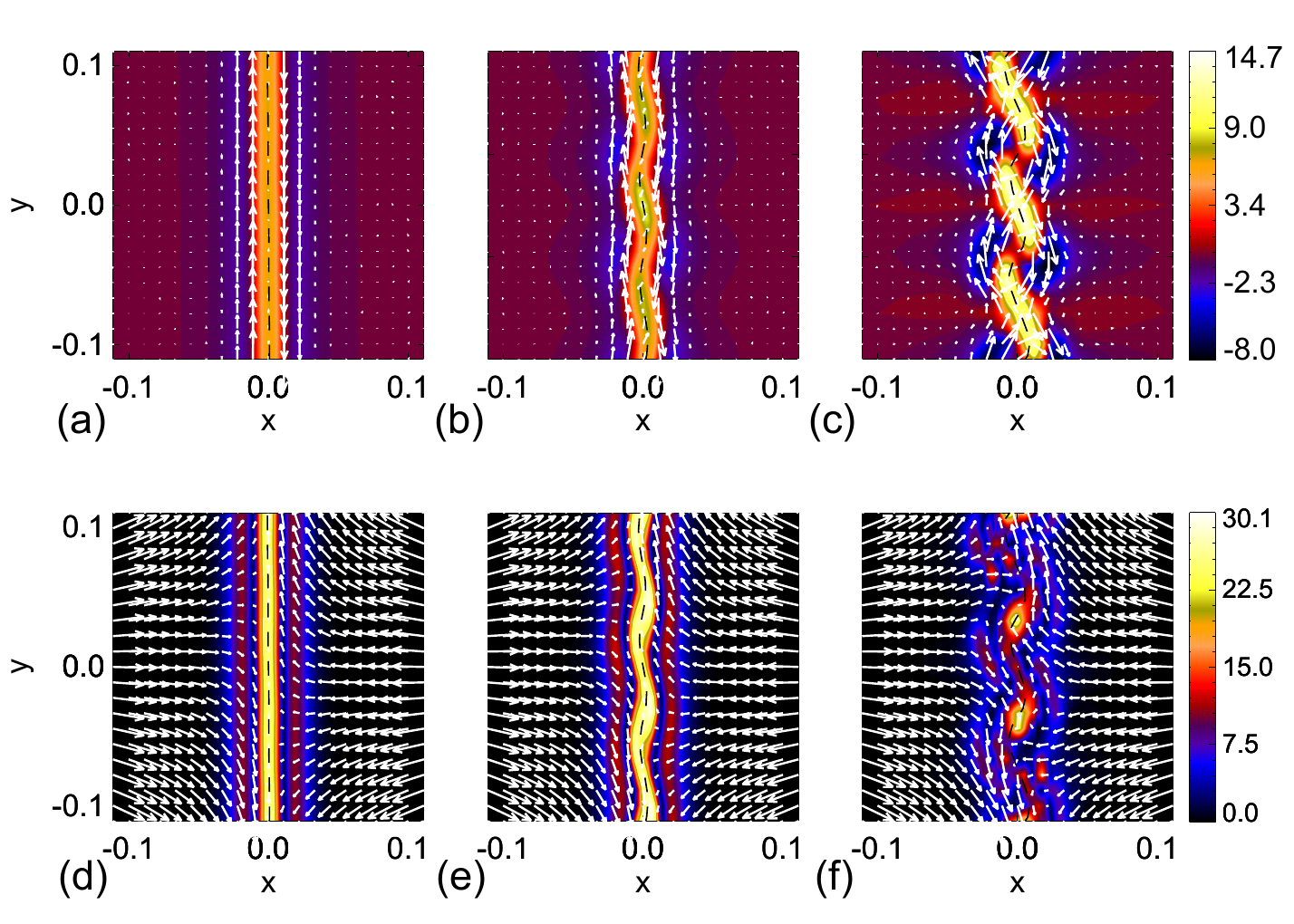}
\caption{Development of the instability in the current-vortex sheet. Top: contours indicate the current out of the plane and vectors show the plasma velocity. Bottom: contours indicate vorticity density and vectors the magnetic field components in the plane. Dashed line shows the fan plane position. Taken in the plane $z=0.85$ at $t=6$ (a,d), $t=7$ (b,e) and $t=8$ (c,f) for simulation 9.}
\label{fig:roll}
\end{SCfigure*}

\begin{SCfigure*}
\centering
\includegraphics{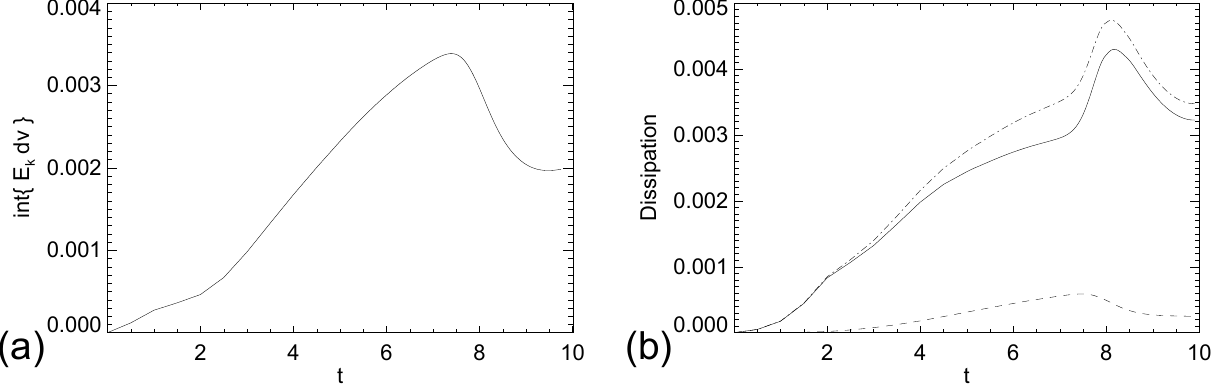}
\caption[Dissipation of kinetic energy in the volume.]{(a) $\int{E_{k} dV}$; kinetic energy integrated over the volume of the box as a function of $t$. (b) Energy dissipation through $\eta J^{2}$ (solid) and $Q_{visc}$ (dashed) and $\eta J^{2}+Q_{visc}$ (dot-dashed) as a function of $t$. Both for case 9 (with $\eta=5\times 10^{-4}$ and $\mu=1\times 10^{-5}$).}
\label{fig:rates}
\end{SCfigure*}

\subsection{Nature and evolution of the sheet breakup}
We now describe the growth and evolution of the KH instability in our current-vortex sheet. In this section we discuss the results of simulation 9, which are representative of the evolution of the instability in the various simulations. Loosely speaking, the instability involves a filamentation of the current layer, with the formation of vortical flows in the plane of the velocity shear. 
Figure \ref{fig:rolltop}(a-c) shows the development of the KH vortices by plotting the associated component of velocity perpendicular to the $x=0$ plane. In Fig.~\ref{fig:avr}(a) this is quantified by plotting the azimuthally averaged value of $\rho v_x^2/2$ at $x=0$ (the original position of the fan plane). From both figures it is evident that the instability initially develops at around $r=0.9$. The vortex tubes form aligned to the radial magnetic field, thus reducing the damping effects of magnetic tension.
Analyzing Fig.~\ref{fig:avr}(b) it is clear that $r=0.9$ coincides with the radial peak in vorticity. Thus, the size of the vortex hole in the fan plane (discussed above) dictates the starting point of the KH instability growth. We note that the development of the instability depletes the net vorticity within the current-vortex layer at later times.

Following the guiding influence of the radial magnetic field the vortex tubes spread outwards and inwards from the initial radius of formation. As they spread outward from the null they encounter a longer (in the $\hat{\bf \theta}$ direction) and thinner (in the $\hat{\bf x}$ direction) shear layer and so branch off in order to maintain a diameter approximately equal to the thickness of the shear layer (Fig. \ref{fig:rolltop}(c)). Conversely, as they spread inwards toward the null they coalesce. Once out of the linear growth phase the KH vortices saturate as they reach the width of the shear layer and a new slowly varying state is reached. 

Figure \ref{fig:roll} shows how the development of the instability affects the structure of the current sheet. The instability results in a rippling or kinking of the shear layer, in the same way as described in Ref.~\onlinecite{Dahlburg1997}. In the strongly sheared stagnation point flow between each vortex a strong current layer forms. This fragments the current sheet into filaments that lie between each of the branched off vortex tubes and appear as fingers of current in Fig.~\ref{fig:rolltop}(d-f). This additional localization of the current layer in the azimuthal direction naturally leads to the formation of twisted magnetic field structures (though Amp\`{e}re's law) along each current filament (Fig.~\ref{fig:roll}(f)). The circular component of these fields is, however, small in comparison with the radial guide field resulting in only a small deformation of the fan plane (black dashed line, Fig.~\ref{fig:roll}). Similar circular magnetic field structures have been observed to form between the plasma vortices in 2D simulations with comparatively better resolution of the shear layer \citep[e.g.][]{Keppens1999, Antognetti2002}. However, in the 2D scenario the magnetic tension of the field in the flow vortex leads to additional circular magnetic fields co-aligned with the vortices which we do not see here, perhaps as a result of the strong transverse field that is present. 

As a result of the formation of the flow vortices, the kinetic energy within the volume is more efficiently converted to local twist within the shear layer and dissipated through ohmic heating than prior to the onset of the instability. Figure \ref{fig:rates} shows this for case 9 as a drop in the volumetric kinetic energy and viscous dissipation along with an increase in ohmic dissipation during the non-linear phase (beginning around $t=7.5$). Considering the total dissipation within the volume it is clear that the onset of the instability has led to an increase in the localized heating of the plasma around the fan plane. 

Once the instability saturates the layer settles down toward a new slowly varying state, in which no secondary instabilities of the current filaments arise but some filaments are seen to begin to coalesce through the `zipping up' of adjacent current branches. It was conjectured by Dahlburg \& Einaudi \cite{Dahlburg2002} and subsequently confirmed by Onofri {\it et al.} \cite{Onofri2004} that a strong guide field stabilizes secondary kinking instabilities leaving a state in which the coalescence instability is the dominant mode, and our results imply that this is still the case when there is a strong shear flow present. Comparing the averaged shear layer widths and strengths from before and after the onset of the instability; the layer has become widened and the shear reduced. Thus, the KH instability acts as a key to the relaxation (on average) of the stress across the fan plane by allowing the system to transition to a state with a steady sharing of flux between the two topologically distinct regions, as discussed below.

\begin{figure*}
\centering
\includegraphics{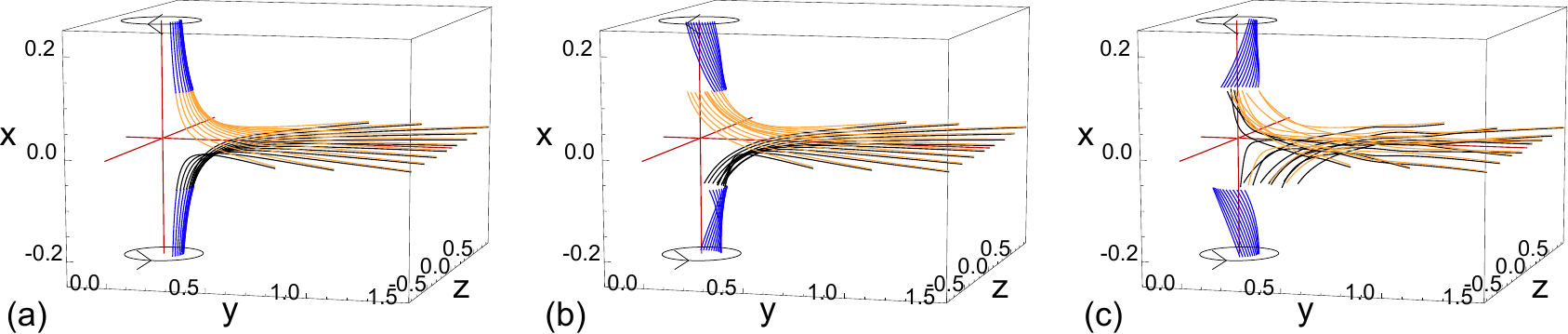}
\caption{A selection of field lines traced from the boundaries of the box (a) initially ($t=0$), (b) following the main current sheet formation ($t=5.0$) and (c) after the onset of the instability ($t=8$). The red field lines depict the position spine and fan of the null and the circular arrows the direction of boundary driving. A description of the connectivity change is given in the text.}
\label{fig:connections}
\end{figure*}

\begin{figure}
\centering
\includegraphics{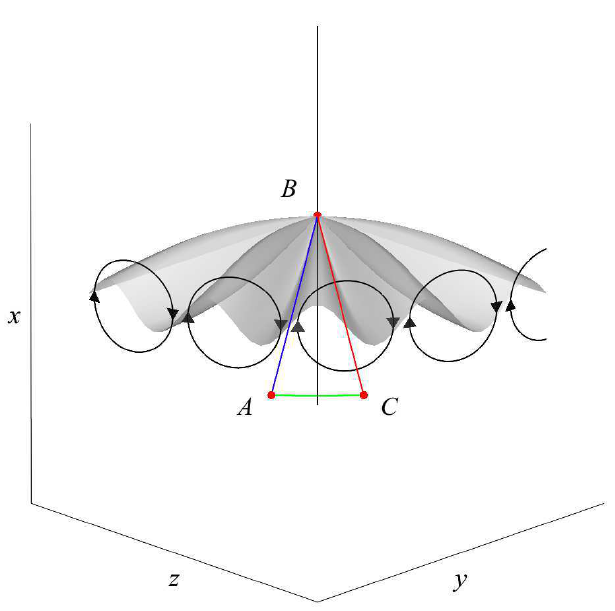}
\caption{Sketch of the paths considered in Eqs. (\ref{patheq1}, \ref{patheq2}). The grey isosurface shows the perturbed fan plane and the black streamlines the plasma flow. The null point is located at B.}
\label{fig:KH-diagram}
\end{figure}

\subsection{Reconnection Rate}
Before going into the details of how to measure the reconnection rate in the layer, it is instructive to consider whether such local variations in connectivity could have any effect on the global connectivity of the field around the null. Figure \ref{fig:connections} shows how connections near the null change through the onset of the KH instability in the layer. Initially the blue field lines (plotted from foot points on the driving boundaries) are connected to the yellow and black field lines (plotted from the foot points on the side boundaries). Once the driving begins the foot points of the blue field lines are advected and, after the main current sheet forms at the fan, these field lines begin to slip on to their nearest neighbors in a circular manner. This slippage is the well studied torsional fan reconnection scenario, during which no flux is transferred across the fan plane. The reconnection rate for this slippage is found from the maximum of the integral of $E_{\parallel}$ along field lines passing very near to the spine and out along the fan.

However, once the instability occurs in the current sheet numerous connection change events occur across the fan plane -- see Fig.~\ref{fig:connections}(c). That is, flux has been exchanged between the two topologically distinct regions. How do we evaluate and interpret the reconnection rate in this case? Clearly, once the KH instability begins, both kinds of reconnection are taking place. Focusing on the connection change across the fan plane it is evident from Fig.~\ref{fig:roll} that for each KH vortex the flux transferred in one direction across the fan is matched by an equal amount transferred in the other direction, so that the net flux transfer is zero. This can be expressed mathematically through the fact that the integral of $\mathbf{v}\times\mathbf{B}$ around a closed curve ($\mathcal{C}$) in the fan plane beyond the edge of the current sheet (i.e.~in the ideal region in which ${\bf E}=-{\bf v}\times{\bf B}$) is zero:
\begin{eqnarray}\label{fluxtransfereq}
\int_{\mathcal{C}}{\mathbf{v}\times\mathbf{B} \cdot d\mathbf{l}}&=&-\int_{\mathcal{C}}{\mathbf{E} \cdot d\mathbf{l}}= -\iint_{\mathcal{S}}{\mathbf{\nabla} \times \mathbf{E} \cdot d\mathbf{S}}\nonumber\\
&=&\iint_\mathcal{S}{\frac{d\mathbf{B}}{dt} \cdot {\bf n}dS}=0,
\end{eqnarray}
where $\mathcal{S}$ with normal vector ${\bf n}$ is the portion of the fan surface bounded by $\mathcal{C}$, and ${\bf n}$ is perpendicular to ${\bf B}$ by definition. The gross rate at which flux is driven across the fan plane by  {\it{all}} of the KH vortices gives the {\it{total}} associated rate of reconnection. To find this we sum the contribution from each half of each vortex tube in turn. We exploit the path independence of $\mathbf{E}$ (see Eq.~(\ref{fluxtransfereq})) and consider the path depicted in Fig. \ref{fig:KH-diagram}. Due to the path independence, the rate at which flux is driven across the fan by half of the depicted vortex tube can be written as
\begin{equation}\label{patheq1}
-\int_{AC}{\mathbf{v}\times\mathbf{B} \cdot d\mathbf{l}}=\int_{AC}{{\bf E} \cdot d\mathbf{l}}=\int_{ABC}{{\bf E} \cdot d\mathbf{l}}
\end{equation}
where $AB$ and $BC$ are field lines lying in the fan surface, and so the rate of change of flux associated with half of one KH fluctuation, $\dot{\Psi}_{KH}$, is given by
\begin{equation}\label{patheq2}
\dot{\Psi}_{KH}=\int_{AB}{E_{\parallel} dl} -\int_{CB}{E_{\parallel} dl}.
\end{equation}
Therefore, the reconnection rate associated with one vortex tube is double the difference between the integral of $E_{\parallel}$ along the strong current lane between two adjacent tubes ($AB$) and the integral of $E_{\parallel}$ through the weak current along the tube center ($CB$). For the total gross rate of flux transfer across the fan plane this must then be summed over all of the vortex tubes.

To evaluate the total reconnection rate in practice, for each snapshot in time, we integrate $E_{\parallel}$ along a large number ($1800\sim 3600$) of field lines in the fan plane. Each integral is evaluated between $r=0.05$ (since field tracing at the null is numerically problematic) and $r=2.8$ (the edge of the boundary damping region). The data is then binned to remove long wavelength modes resulting from the Cartesian grid, after which the difference between peaks and troughs is summed over the entire angular distribution. This gives an approximation to the total rate of flux transfer across the fan plane
\begin{equation}
R_T=\left(\dot{\Psi}_{KH}\right)_{total}. \label{KHrr}
\end{equation} 

Consider now the rate of rotational slippage occurring in the volume associated with the torsional fan reconnection that occurs even in the absence of the instability. To evaluate this we take advantage of the cylindrical symmetry of the driving. We know that despite the small scale fluctuations in the main fan plane current sheet the driven foot points are still smoothly rotating and thus, on average, the flux must change connections in a symmetric manner to preserve the overall symmetry. Therefore, we define the rate of rotational slippage of flux as the maximum of the azimuthal average of the integral of $E_{\parallel}$, i.e.
\begin{equation}
\dot{\Psi}_{s}=\left(\left\langle\int{E_{\parallel} dl}\right\rangle\right)_{max},
\end{equation}
where $<..>$ denotes an average over the azimuthal angle. In practice this occurs along field lines lying asymptotically close to the spine and fan. Note that this defines an overall reconnection rate, including rotational slippage both within the fan current sheet, and the large-scale current concentrations around the spine. Using these definitions we can quantify how flux is reconnected as various parameters of the system are varied.

\begin{figure*}
\centering
\includegraphics{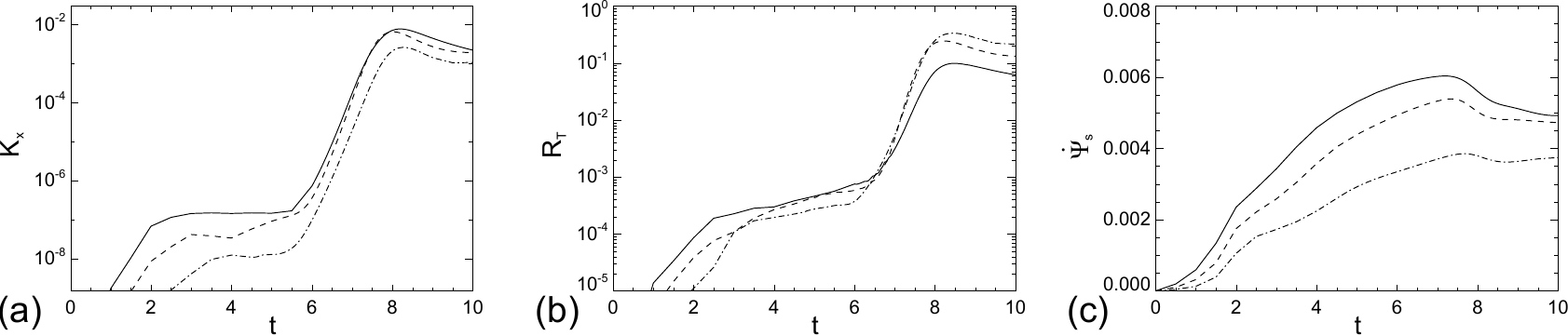}
\caption{Log plots of $K_{x}$ (a), $R_{T}$ (b) and $\dot{\Psi}_{s}$ (c) as a function of $t$. For simulation runs with $\mu=1\times 10^{-5}$ and $\eta=2\times 10^{-4}$ (dot-dashed), $5\times 10^{-4}$ (dashed) and $1\times 10^{-3}$ (solid).}
\label{fig:timeevo}
\end{figure*}

In general, in each simulation, $\dot{\Psi}_{s}$ grows steadily under the action of the continued boundary driving until, at some point, the KH instability fragments the the current-vortex layer, decreasing $\dot{\Psi}_{s}$ as $R_{T}$ grows rapidly. Figure~\ref{fig:timeevo}(b-c) shows this for cases 8-10 where it is clear that $R_{T}$ rapidly grows to dominate over $\dot{\Psi}_{s}$ as a result of the recursive nature of the KH induced connection change and the large number of reconnection sites in the fan plane.

\begin{figure}
\centering
\includegraphics{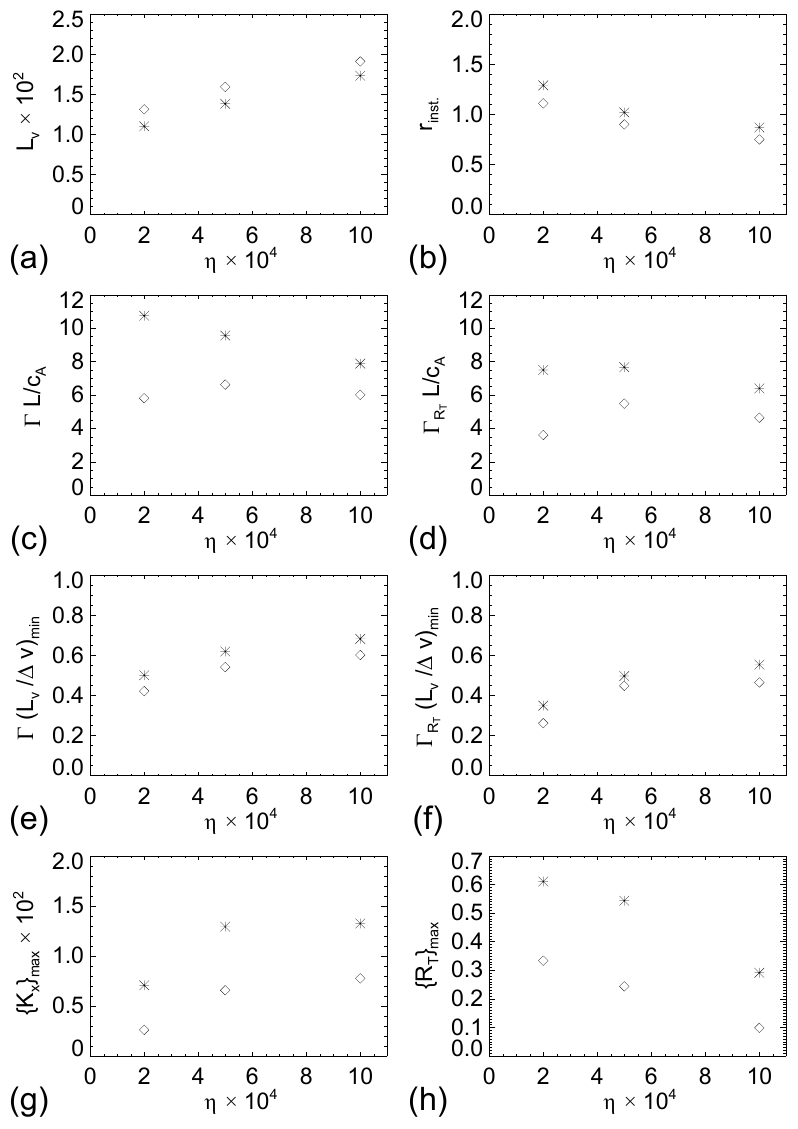}
\caption{Scalings with $\eta$ of the (a) average shear layer width ($L_{v}$, between $r=0.6$ and $2.8$), (b) radius of initial instability growth ($r_{inst.}$), (c,d) growth rates of ($K_{x}$, $R_{T}$) (($\Gamma$, $\Gamma_{R_{T}}$), normalized by $L/c_{A}$), (e,f) growth rates ($K_{x}$, $R_{T}$) (normalized by $(L/\Delta v)_{min}$), (g,h) peak values of ($K_{x}$, $R_{T}$). Asterisks: with numerical viscosity, diamonds: $\mu=1\times 10^{-5}$.}
\label{fig:etavs}
\end{figure}

\begin{figure}
\centering
\includegraphics{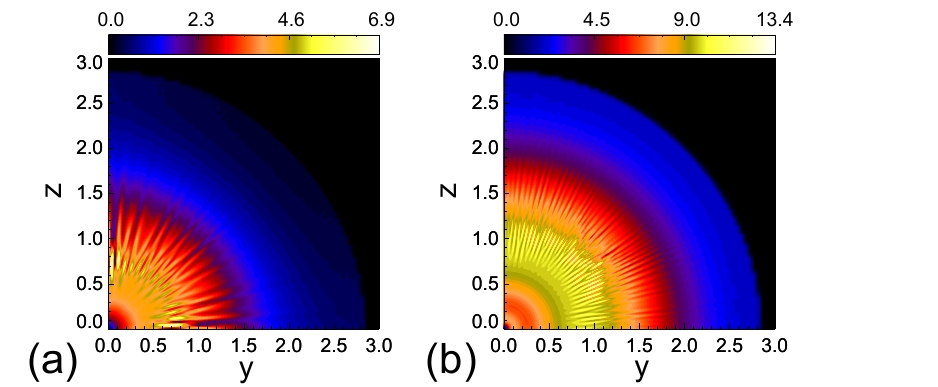}
\caption{Current density in the $x=0$ plane for (a) case 13 (at $t=6.5$) and (b) case 11 (at $t=7.5$).}
\label{fig:rolleta}
\end{figure}

\begin{SCfigure*}
\centering
\includegraphics{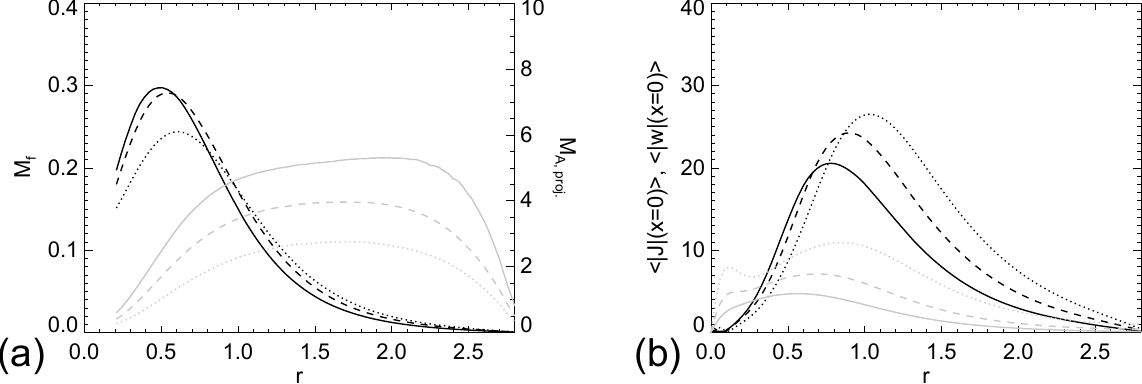}
\caption{(a) Mach numbers for different $\eta$ ($M_{f}$: black, $M_{A,\;proj.}$: grey). (b) The changing vorticity (black) and current (grey) densities with $\eta$. For cases 8 (dotted), 9 (dashed) and 10 (solid) when $t=4$ (prior to the instability).}
\label{fig:compmac}
\end{SCfigure*}

\subsection{Quantitative properties of the system}
Here we explore briefly how the behavior described above changes as we vary the resistivity and viscosity of the plasma. As the resistivity of astrophysical plasmas such as the solar corona is incredibly small (and therefore impossible to simulate numerically) it is important that we understand how any simulated reconnecting system behaves as the resistivity is reduced. Typically, this involves finding scaling laws for parameters such as the growth rate of the instability or the peak reconnection rate. A full scaling analysis is impractical here given the size of the data sets and computational power required, but we have made a preliminary investigation.

\subsubsection{Quantitative properties of the current-vortex sheet}
We first consider how the properties of the current-vortex sheet changes as we vary $\eta$. As $\eta$ is reduced the current layer, and thus the velocity shear layer, becomes thinner. This can be seen in Fig.~\ref{fig:etavs}(a) which shows the mean velocity layer thickness ($L_{v}$) between $0.6 \leq r \leq 2.8$ in cases 8 to 13 at $t=4$. As the shear layer thins, the vorticity hole becomes wider (due to the hyperbolic shape of the field) and the KH instability sets in initially at a larger radius (Fig. \ref{fig:etavs}(b)). This suggests that at realistic coronal parameters the instability of the sheet will occur at large distances from the null, but through the spreading of its influence described in the previous section will still dominate a significant portion of the fan plane current sheet. In competition with the above effect, the increase in viscosity between cases 8 to 10 and 11 to 13 clearly widens the shear layer so that in more viscous fluids the vortices form closer to the null. 

An important consequence of the formation of a thinner shear layer is that the number of vortices that form greatly increases  (Fig. \ref{fig:rolleta}). This suggests that as $\eta$ is reduced and the layer thins the associated gross flux transfer across the fan plane may increase. We will explore this further below. Lastly, at large values of $\eta$ the magnetic field may slip through the plasma more readily than at lower values. Thus, as $\eta$ is lowered, so the relative velocity to magnetic shear ($M_{A,\; proj.}$) decreases. This is shown in Fig. \ref{fig:compmac} which also shows the increasing current and vorticity density as a result of the thinning of the layer.

\subsubsection{Quantitative properties of the instability}
We now consider the dependence of the instability on $\eta$ and $\nu$. Note that for a KH-type instability we would expect no dependence on $\eta$ in the linear phase. However, there is an indirect influence due to the varying properties of the shear layer that forms, as discussed above. Moreover, the properties of the instability as it saturates can be expected to depend on the dissipation parameters, as discussed below.

The evolution and growth of the instability of the current-vortex layer can be quantified by integrating the kinetic energy associated with the component of the flow perpendicular to the $x=0$ plane, i.e. 
\begin{equation}
K_x=\iint{\frac{1}{2}\rho v_{x}^{2} dydz},
\end{equation}
evaluated in the $x=0$ plane. This can be compared with the gross rate of flux transfer across the fan surface ($R_{T}$: Eq. \ref{KHrr}). As shown in Fig.~\ref{fig:timeevo}, both quantities exhibit an exponential growth phase during the early stages of the instability. The increase in $R_T$ lags behind the growth of $K_x$ as expected for a KH-type instability since the linear phase of the instability is ideal, and initially the kinetic energy of each vortex is expended in ideally deforming the fan surface (dashed black line, Fig.~\ref{fig:roll}). By assuming that the growth phase approximates $\sim e^{\Gamma t}$ and normalizing against the Alfv\'{e}n travel time across the width of the simulation box (using the Alfv\'{e}n speed at the spine foot points and the distance between the $x$-boundaries: $t_{A}\approx L/c_{A}(x=0.25)=0.5/0.5=1$) we can compare the growth rates of $K_x$ and $R_T$ (denoted $\Gamma$ and $\Gamma_{R_{T}}$, respectively) relative to a typical time scale for the whole system. It should be noted, however, that as these growth rates are global quantities they inherently include many different effects that can affect the growth rate of the KH instability (such as different dominant wave numbers, shear layer widths, Alfv\'{e}n Mach numbers and magnetic and plasma Reynolds numbers of each shear layer).

Fig. \ref{fig:etavs}(c) shows the growth rate of $K_{x}$ for the different cases that were studied. In general, the growth rate of the instability is seen to increase as $\eta$ is reduced. The exception to this is case 8 (with $\mu=1\times10^{-5}$ and $\eta=2\times10^{-4}$), where the wavelength of the dominant wave-mode has likely become comparable to the viscous length scale. The growth rates of $R_{T}$ (Fig. \ref{fig:etavs}(d)) follow a similar trend, but at the lowest values of $\eta$ are reduced as a likely result of increased numerical diffusion leading to an underestimation of the value of $\eta$. 
 
One of the factors affecting the growth of the instability that we can normalize against is the changing thickness of the shear layer. Specifically, we can consider the effect of normalizing against the minimum travel time across the vorticity layer ($(L_{v}/\Delta v)_{min}$), taken before the onset of the instability (at $t=4$). Fig.~\ref{fig:etavs}(e,f) shows the same growth rates as above but with the effect of the thinning shear layer removed. In this case the growth rates of both quantities now decrease with decreasing $\eta$. This shows that the dominant factor in the increase in the global growth of the KH instability as $\eta$ is reduced is the thinning of the shear layer. What is clear from our results is that the growth rates of both $K_x$ and $R_T$ are linked and that the relative strength of resistivity to viscosity (i.e. the magnetic Prandtl number) plays an important role in setting up the initial layer, and thus how fast the KH instability of the layer grows.

Lastly, we note that small-scale near-turbulent reconnection events around the edge of the flow vortices have been seen in 2D simulations (e.g. Ref.~\onlinecite{Keppens1999}) to help drive the plasma circulation and increase the growth rate. This effect is beyond what is currently available to study at the resolution of these simulations but could also play an important role in the dynamics of the current-vortex sheet. 

Once the vortices grow to be comparable to the width of the shear layer they saturate and reduce in amplitude. At this point both $K_x$ and $R_T$ peak, before reaching a new steady level. Fig. \ref{fig:etavs}(g,h) shows the peak values attained. There appears to be a reduction in the peak value of $K_x$ with decreasing $\eta$, although this does not appear systematic and may be partly due to an under-resolution of the vortex structures in $v_x$ at the smallest values of $\eta$. It could also be that the maximum energy content of the perturbation flow associated with the vortices is reduced as the non-linear phase is entered earlier with a reduction in the width of the shear layer. By contrast, the peak reconnection rate $R_T$ increases approximately linearly as $\eta$ is reduced. This surprising result arises because as $\eta$ is reduced there are many extra vortex tubes produced as the shear layer thins.

\section{Discussion}
In this work we investigated first the self consistent formation, and then the stability to Kelvin-Helmholtz-type vortices, of the current-vortex sheet which forms in response to twisting motions around the spines of a symmetric 3D magnetic null point. The formation of the current-vortex layer was found to be complex, with local non-ideal effects brought about by strong current regions that form along the spine lines. This strongly affects the initial region of KH growth when the layer is KH unstable.

The instability leads to the breakup of the planar current layer at the fan, into vortex structures in the plane of the flow and magnetic shears, consistent with earlier studies of 2- and 2.5-dimensional current-vortex sheets. Due to a combination of the field geometry and resistive and viscous diffusion, the unstable region with high magnetic and flow shears is confined at intermediate radii from the null point. The widths and magnitudes of the current and vorticity shear layers are consistent with a KH-type instability according to the previous theory \citep{Einaudi1986}. There are a number of aspects of our simulations that point to the observed instability being of a KH-type. First, the growth of the reconnected flux ($R_T$) lags behind the growth of the kinetic energy  of the perturbation, suggesting that in the early stages the instability is predominantly an ideal one. Furthermore, the instability is associated with a rippling or kinking of the shear layer, consistent with the KH-dominated regime in 2D as discussed in Ref.~\onlinecite{Dahlburg1997}. One new effect that we observed was a `branching' of the vortex tubes / current filaments in the direction transverse to the shear. This is an effect that arises due to the fully 3D nature of the system, in that the layer gets longer (in the azimuthal direction of the shear flow) and thinner as we move away from the null in this transverse direction.

Once formed, it was found that the instability quickly grows to dominate the majority of the current-vortex sheet. The strong current lanes between the `branched' vortex tubes more efficiently heat the plasma in the layer and enable a rapid recursive sharing of magnetic connections between the two previously separate topological regions. This associated connectivity change dominates over the torsional slippage associated with the global driving flow. This shows how, subject to a smooth deformation, instabilities in the fan surface current layer of 3D null points can lead to a sudden burst of mixing of plasma between the previously distinct regions on either side of the separatrix surface. Such recursive reconnection is also a feature of reconnection between nulls connected by multiple separators \citep{Parnell2010}. Plasma heating at fan separatrix surfaces has previously been proposed to be important in the solar corona \citep[e.g.][]{priest2005}. Our results suggest that this heating could be significantly enhanced by instabilities of the current vortex-sheet that forms at the separatrix. It should be noted that the effect of thermal conduction in these simulations has been neglected. This would act to transport thermal energy away from the current layer along field lines, and thus reduce the temperature, altering the pressure/density distribution within the current-vortex layer. While we do not expect that our qualitative results would change, we might expect the region of the fan plane that is unstable, and the growth rate, to be slightly modified. The effect of additional terms in the energy equation should be considered in future studies.

As the instability of the current-vortex sheet is strongly dependent on the driving and plasma parameters we have only focused on a small parameter range to best optimize the available resolution. We focused our attention on the importance of viscosity and resistivity in the dynamics of the layer. We were hampered by the difficulty in resolving the fine scale structure following the formation of the layer. However, what became clear is that both the resistivity and the viscosity of the plasma have a strong effect on the initial position, growth rate and saturation level of the KH instability in the fan plane current-vortex sheet. This is perhaps counter-intuitive since the KH instability is an ideal one, but the key point is that the properties of the shear layer itself are strongly dependent on $\eta$ and $\nu$, which means that the growth rate of the instability also is, albeit indirectly. In addition, changing other parameters would also affect the layer. For instance in 2D it is known that increasing the compressibility of the plasma or the driving speed so that $M_{f} \gtrsim 1.3$ produces shocks where the plasma is accelerated by the vortices \citep{Miura1984,Shen2000}. 

We note that for continued driving beyond the end of these simulations the kink instability may set in around the spine lines which would destroy the formation of the current-vortex shear layer, leading to patchy reconnection across the fan plane \citep{Pariat2009}. 
We would expect that the kink instability sets in sooner for systems which are more strongly driven or less resistive, as more twist may build up around the spine. This suggests that there may only be a window in time for which the KH instability forms in the fan plane current-vortex sheet before the kink instability destroys the shear layer. It remains to be seen if at more realistic coronal parameters this window is too short for the KH instability, under these driving conditions, to be a realistic means for sudden energy release.

\section{Conclusions}
In future there are a number of extensions to our study that could be considered. Relaxing the symmetry of either the initial magnetic null point field or the driving flow is likely to profoundly affect the locations in which the current-vortex sheet is unstable. In addition, we have so far not probed the range of parameters appropriate to a tearing-type instability. Furthermore, we have observed a similar type of instability when the null is subjected to shearing rather than the highly-symmetric rotational perturbations considered here. The resulting spine-fan reconnection mode is arguably more generic, and thus it may be a good candidate for rapid energy release through either the tearing or KH instabilities. This will form the basis of a future study. The development of a more turbulent state should also be studied, though this requires much higher resolution of the current-vortex sheet, and a different numerical approach may be necessary. The instability of current layers around 3D null points will also have a profound effect on the acceleration of particles in these locations, with the well-collimated beams observed in existing studies \citep[e.g.][]{stanier2012} likely to be modified.

We conclude that the fragmentation via the KH instability of the torsional fan current-vortex sheet could provide a rapid mechanism for energy release and connectivity change between the two topologically distinct regions separated by the fan separatrix surface. However, the setup considered here is highly symmetric and other instabilities may be triggered first if wider parameter spaces are considered. Further work is needed to determine if this mechanism is a realistic candidate for sudden energy release in parameter spaces typical of astrophysical plasmas. We hope that the techniques and understanding developed in this work could be used to probe this scenario more deeply in the future.

\begin{acknowledgments}
We would like to thank Rekha Jain and K. Galsgaard for many useful discussions and suggestions and additionally we thank K. Galsgaard the use of his 3D resistive MHD code in this study. We also wish to thank the anonymous referee for their useful suggestions. Computational time was given by the MHD cluster of the University of St. Andrews. PW and DP acknowledge the financial support of EPSRC and The Leverhulme Trust, respectively.
\end{acknowledgments}

\bibliographystyle{aipnum4-1} 
\bibliography{biblionew}

\end{document}